\documentclass[aps,pra,floatfix,showpacs,preprint,tightenlines,superscriptaddress]{revtex4}
\usepackage{graphicx}
\usepackage{epsf}
\begin{document}
\draft
\title{Lorentz invariant intrinsic decoherence.}
\author{ G.~J.~Milburn}
\address{Centre for Quantum Computer Technology,\\
Deprartment of Physics, School of Physical Science,\\  The University of
Queensland, Queensland 4072 Australia.}
\date{\today}

\begin{abstract}
Quantum decoherence can arise due to classical fluctuations in the parameters which define the dynamics of the system.   In this case  decoherence, and complementary noise, is manifest when data from repeated measurement trials are combined. Recently a number of authors have suggested that fluctuations in the space-time metric arising from quantum gravity effects would correspond to a source of intrinsic noise, which would necessarily be accompanied by intrinsic decoherence. This work extends a previous heuristic modification of Schr\"{o}dinger dynamics based on discrete time intervals with an intrinsic uncertainty. The extension uses unital semigroup representations of space and time translations rather than the more usual unitary representation, and does the least violence to physically important invariance principles. Physical consequences include a modification of the uncertainty principle and a modification of field dispersion relations, in a way consistent with other modifications suggested by quantum gravity and string theory . \end{abstract}

\date{\today}
\maketitle
\section{Introduction}
The precision with which intervals of time and length can be measured
is limited by intrinsic quantum uncertainties\cite{BCM96}. 
The limit on precision is determined by 
fundamental constraints on estimating the parameter of an appropriate time or space translation.
These limits arise from the statistical distinguishability of quantum states and 
reflect the geometry of Hilbert space itself.  

In practice however, precision is limited by 
interactions between the measured system and other degrees of freedom (the environment) over which we have little control. Such interactions limit precision as noise is added to the measurement outcomes reflecting our lack of knowledge of the precise state of the environment. The flip side to added noise is decoherence: the interactions with the environment destroy coherence between superposed quantum states in some specific basis. Studies of environmentally induced decoherence over the last three decades have given a reasonably good picture of the process \cite{Leggett,Paz}  although detailed comparison to experiment is relatively recent. Quantum decoherence can also arise due to classical fluctuations in the parameters which define the dynamics of the system.   Fluctuations in the space-time metric would correspond to a source of intrinsic noise, which would necessarily be accompanied by intrinsic decoherence\cite{Kok2003}. 

Decoherence is often invoked  to explain the lack of quantum effects in macroscopic systems.  While this is often  the case for environment induced decoherence, a number of authors\cite{GRW,Milburn91,Percival,Adler} have speculated that an intrinsic decoherence may exist to establish classical behaviour at some level. In this paper we extend a previous approach based on stochastic time \cite{Milburn91} to stochastic space. This provides a path to a  Lorentz invariant field theoretical formulation of a model of intrinsic decoherence. 

These heuristic modifications of the Schr\"{o}dinger equation should more properly be viewed in a like manner to environmental decoherence, but in which  the quantum nature of the environment is left unspecified.   Recently Gambini et al.\cite{Gambini},  have shown that a recent proposal for quantisation of gravity based on discrete space-time is consistent with the model of intrinsic decoherence discussed in \cite{Milburn91}.  The extension proposed in this paper like wise has consequences that have been previously considered in the context of quantum gravity.   Intrinsic decoherence due to spatial displacements leads to a modification of the uncertainty principle which is of the same general form as considered in the context of quantum gravity by a number of authors\cite{Ahluwalia,Kempf,Capozziello}. When extended to the relativistic case, in particular the electromagnetic field,  we find that the dispersion relation for the free field must be modified. A similar effect has also been suggested for models of quantum space-time\cite{GambiniPullin} 

Space and time parameterise fundamental symmetry groups. 
The action of a group element on a physical state is  represented by 
a unitary operator on Hilbert space. Conventionally we consider continuous 
representations of these symmetries which reflect the strong classical 
intuition that space and time are continuous parameters. In this case
we can define the unitary representations through their infinitesimal action.
The unitary representation is then defined in terms of a hermitian operator which
is the generator of the group. In the case of time translations, the generator is 
the Hamiltonian operator while in the case of spatial translation the generator is the momentum
operator. In a relativistic theory these operators are constructed from the quantum
fields that define the physical systems under investigation.  As Anandan\cite{Anadan2000} has emphasised,
the existence of such unitary representations for space and time translation for any
physical system lends a degree of universality to space and time translation,
universality that is the core of our understanding of spacetime itself.

Quantum mechanics is an irreducibly statistical theory. The states of a physical system 
in quantum theory enable one to construct probability measures for the outcomes of 
all possible measurements made upon the system.  Gleason's theorem tells us 
that the most general way to generate these probability measures in 
the Hilbert space formulation of quantum mechanics is in terms of a positive trace class operator
the density operator $\rho$.  In the nonrelativistic formulation two rules, one kinematical and
one dynamical, are required to specify how probability amplitudes change in space and time. These rules were first given
implicitly by Schr\"{o}dinger in terms of a partial differential equation for a complex valued function of space and time. 
However they can be stated in terms of unitary representations of the space and time translation groups. 
Space and time translations then correspond to rotations of vectors in Hilbert space. 

Spatial and temporal translations are then represented by the 
one parameter unitary transformations,
\begin{eqnarray}
\rho(X) & = & e^{-iX\hat{p}/\hbar}\rho_0e^{iX\hat{p}/\hbar}\\
\rho(t) & = & e^{-it\hat{H}/\hbar}\rho_0e^{it\hat{H}/\hbar}
\end{eqnarray}
or in differential form,
\begin{eqnarray}
\frac{d\rho(X)}{dX} & = & -\frac{i}{\hbar}[\hat{p},\rho(X)]\\
\frac{d\rho(t)}{dt} & = & -\frac{i}{\hbar}[\hat{H},\rho(t)]
\end{eqnarray}
The representation 
of the group of spatial and temporal translations in terms of the unitary operators generated by energy and momentum
of course leads directly to the conservation of energy and momentum. The Schr\"{o}dinger prescription 
thus reflects how probability amplitudes respect our most important symmetry principles. 

Let us now take a different perspective. We will take rotations of vectors in Hilbert 
space as fundamental. When we make measurements on the
corresponding physical systems, these rotations will appear as spatial or temporal translations.
The space and time parameters thus inferred are then to be regarded as macroscopically determined parameters that help
us coordinate the results of measurements made under different experimental conditions. 

Within this perspective let us now ask the following question: how can we change the Schr\"{o}dinger rules
causing the least violence to energy and momentum conservation ?  To answer this we need to 
consider in a little more detail the kinds of experiments that enable us to estimate space and time 
translations.   The high precision measurement of space and time intervals are in fact determinations of 
the statistical distinguishably of quantum states through repeated preparation and measurement. 

\subsection{Measurement of time intervals.}
\label{standard_time}
How do we make a quantum clock ? Answer: superpose two sates of definite energy. Indeed this is exactly 
how time is currently measured with atomic clocks. 
Let the system be prepared, at time $t=0$, in the state 
\begin{equation}
|\psi\rangle=\frac{1}{\sqrt{2}}(|E_1\rangle+|E_2\rangle)
\end{equation}
where $|E_i\rangle$ are energy eigenstates. The variance of the energy in  this state is $\langle\Delta\hat{H}^2\rangle_0=\Delta_E^2/4$ with $\Delta_E=E_2-E_1$ 

 After a time $t$, in accordance with the dynamical Schr\"{o}dinger rule, the state becomes
\begin{equation}
 |\psi\rangle=\frac{1}{\sqrt{2}}(e^{-i\omega_1 t}|E_1\rangle+e^{-i\omega_2 t}|E_2\rangle)
\label{time_state}
\end{equation}
where $\omega_i=E_i/\hbar$. 

The next step is to measure some quantity represented by an operator that does not
commute with the Hamiltonian.  The simplest choice is the projection operator $\hat{P}_+=|+\rangle\langle+|$ onto the state $|+\rangle=\frac{1}{\sqrt{2}}(|E_1\rangle+|E_2\rangle)$. There are two possible values, $x=0,1$ for the measurement result, with probability distribution
\begin{equation}
p_1(t)=1-p_0(t)=\cos^2\left (\frac{\Delta_E t}{2\hbar}\right ) 
\label{clock-meas}
\end{equation} 

There are two ways this system may be used as a clock. Both cases require us to 
sample the probability distribution in Eq.(\ref{clock-meas}) and thus  both require that we 
prepare a large number of identical systems in the manner 
just described and simultaneously measure the quantity $\hat{P}_+$ on all of them. 

The first and most direct method
is simply to measure the quantity $\hat{P}_+$ and thus infer the probability $p_x(t)$ and thus infer $t$. 
Of course this inference must come with some error which  can easily be determined. There is no escaping this fact
for quantum clocks. The second way is to note that the parameter $\Delta_E/\hbar$ is a frequency. It may be possible to 
change this with reference to a given frequency standard, such as a laser, and then use the sampling of the 
ensemble of systems to keep this frequency locked on a particular value, eg by using a feedback control
to ensure that measurements on the ensemble always tend to give the same value $x$. As in the first method, 
this is necessarily accompanied by some error in a quantum world. 

It is now a simple matter to estimate the uncertainty with which we can infer the parameter $t$.  It suffices to measure the quantity $\hat{P}_+=|+\rangle\langle +|$ on the state given in Eq.(\ref{time_state}).  The average value of this quantity is the probability $p_1(t)$. The uncertainty is this measurement is 
\begin{equation}
\Delta p_1=\sqrt{p_1(1-p_1)}
\end{equation}
The uncertainty in the inference of  the time parameter $t$, is then given by \cite{Wooters81}
\begin{equation}
\delta t=\left |\frac{dp_1(t)}{dt}\right |^{-1}\Delta p_1
\end{equation}
Thus we find the well known result
\begin{equation}
\delta t =\frac{\hbar}{|\Delta_E|}
\label{standard_time}
\end{equation}
Noting that the variance of the energy for the fiducial state is $\langle\Delta\hat{H}^2\rangle_0=\Delta_E^2/4$, we see that
the quality of the inference varies as the inverse of  the energy uncertainty. This is the standard result  for a parameter  based uncertainty principle\cite{BCM96}

To summarise, time measurement in a quantum world requires us to sample a probability distribution by making
measurements on an ensemble of identically prepared systems. A single system and a single measurement won't do. 

\subsection{Measurement of space intervals.}
\label{standard_space}
How do we make a quantum ruler ? Answer: make a standing wave for the position probability amplitude
for a free particle. As we shall see, this will entail doing something very similar to 
the previous discussion for measuring time.  A standing wave 
for the position probability amplitude requires a superposition of momentum eigenstates of equal and 
opposite momentum,
\begin{equation}
|\psi\rangle=\frac{1}{\sqrt2}(|p_1\rangle+|p_2\rangle)
\end{equation} 
This state is an energy eigenstate of a free particle, and the 
resulting quantum ruler will not change in time. However it is not an eigenstate of momentum, the momentum uncertainty is given by $\langle\hat{p}^2\rangle_0=\Delta_p^2/4$ where $\Delta_p=p_2-p_1$. 

 In writing this state with the particular choice of 
relative phase between the two components, we are assuming 
that the origin is located at an antinode of the standing wave.
We now translate the ruler (ie. 'pick up' the ruler and move it...that is how one uses a metre stick !) 
to a new position, labelled $X$. The state now changes 
in accordance with the kinematical Schr\"{o}edinger rule as
\begin{equation}
|\psi(X)\rangle=e^{-ik_1X}|p_1\rangle+e^{-ik_2X}|p_2\rangle
\end{equation}
where $k=p/\hbar$. We now measure some quantity on this state, which in analogy with the time example 
in the previous section, we choose to be conjugate to momentum, that is to say we measure position.
The probability distribution to get a particular result, $x$, is then
\begin{equation}
P(x)\propto\cos^2\left (\frac{\Delta_p(X-x)}{2\hbar}\right )
\end{equation}

To determine how much we had translated the ruler we need to sample this distribution. Again we need
to consider making simultaneous measurements on an ensemble of identically prepared systems, (or we could simply 
repeat the sequence of preparation, translation and measurement steps.  Of course, 
the determination of $X$ will be accompanied by some necessary error of a quantum origin. This is roughly the distance between two successive minima of the probability density, $P(x$).  In this case
one easily sees that this is a consequence of the Heisenberg uncertainty principle resulting from
the initial momentum uncertainty of the state in Eq.(14)\cite{BCM96}.  The uncertainty in the inferred value of $X$ is given by 
\begin{equation}
\delta X\geq\frac{\hbar}{|\Delta_p|}
\end{equation}

We have thus seen that determination of temporal duration and space translation require us to 
sample a probability distribution by making measurements on an ensemble. This leads to 
the well known intrinsic quantum uncertainty limits for time 
and position parameter estimation\cite{BCM96}. The process of preparing the 
ensemble (either by trying to prepare a large number of identical systems, or a process of preparation and 
re-preparation of a single system) is in practice subject to additional sources of error.
Furthermore the measurements are not always
perfect and noise may be added from trial to trial. For this reason the 
actual state used to describe an ensemble  may not 
necessarily be simply a product of identical pure states, but may rather be a density operator reflecting
some additional degree of averaging over unknown sources of noise and error. 
The fact that space and time parameters must be inferred by sampling a 
probability distribution over an ensemble is an 
important insight and opens up an additional path for 
developing a theory of intrinsic decoherence. 

\section{Towards intrinsic decoherence.}
We now modify the unitary representation of space and time translation by 
using semigroup representations. We first recall  that, in estimating a parameter,
multiple trials must be performed and in practice it may not be possible to ensure that 
each trial is identical. Suppose now that {\em in principle} it is impossible for each trial to be 
identical, for some fundamental reason. In that case parameter estimation would necessarily 
be based on an ensemble rather than a pure state. This is the starting point for
intrinsic decoherence. From trial to trial, the experimentalist
seeks to use the same parameterised state. However suppose that the actual
state is different from one trial to the next due to fluctuations in the actual Hilbert space rotations
corresponding to a spatial or temporal shift. The overall parameter estimation problem must then necessarily
be based on a non pure density operator. 

To be more specific we suppose that for each unitary representation of the
parameter transformation there is a minimum Hilbert space rotation angle, $\epsilon$,
and further that the {\em number} of such rotations, from one trial to the next, can fluctuate.
Let $p_n(\theta,\epsilon)$ be the probability that there are $n$ such 
phase shifts for a change in the macroscopic
parameter from $0$ to $\theta$.  We obtain different models for each choice of the probability function
$p_n(\theta,\epsilon)$. Thus the state $\rho(\theta)$  may be written
\begin{equation}
\rho_\epsilon(\theta)=\sum_{n=0}^\infty p_n(\theta,\epsilon)e^{-in\epsilon\hat{g}/\hbar}\rho(0)e^{in\epsilon\hat{g}/\hbar}
\label{solution}
\end{equation}
where $\hat{g}\to\hat{p}$ for spatial translations and $\epsilon\to\mu$ with units of length,  while  $\hat{g}\to\hat{H}$ for temporal translations, and $\epsilon\to\nu$ with units of time. These assumptions are equivalent to assuming that space and time are discrete with fundamental scales determined by $\nu,\mu$. 
 
We also require that, in some limit, the standard unitary  representation is obtained. To this end 
we require
\begin{equation}
\lim_{\epsilon \rightarrow 0} \rho_\epsilon(\theta)=\rho(\theta)=e^{-i\hat{g}\theta/\hbar}\rho(0)e^{i\hat{g}\theta/\hbar}
\label{Schro-limit}
\end{equation}
This condition imposes a restriction on the permissible forms of $p_n(\theta,\epsilon)$.

It is easiest to define the semigroup in terms of its infinitesimal generator. There is 
great deal of freedom in how we do this corresponding to different choices for
the probability distribution for the number of phase shifts.  However we stipulate that it 
must respect the conservation of energy and momentum.  We will use the differential form of the 
parameter transformation
\begin{equation}
\frac{d\rho(\theta)}{d\theta}={\cal D}[\hat{G}]\rho(\theta)
\label{mas_eq}
\end{equation}
where $\theta$ is the parameter, ${\cal D}[\hat{G}] $ is the generator  of a completely positive semigroup
map defined by ${\cal D}[\hat{G}]\rho=\hat{G}\rho\hat{G}^\dagger -\frac{1}{2}[\hat{G}^\dagger \hat{G}\rho+\rho\hat{G}^\dagger \hat{G}]$.
We require that for spatial translations, with generator $\hat{S}$,  ${\cal D}[\hat{S}]\hat{p}=0$ 
while for temporal translations, with generator $\hat{T}$,  ${\cal D}[\hat{T}]\hat{H}=0$. This ensures that momentum and energy are conserved in the 
semigroup transformation.   As a specific example we will take each generator to be a unitary operator of the form;
\begin{equation}
\hat{G}=\epsilon^{-1/2}e^{-i\frac{\hat{g}\epsilon}{\hbar}}
\label{unital}
\end{equation}
which we shall refer to as the {\em unital } case.  Substituting Eq.(\ref{unital}) into Eq.(\ref{mas_eq})  indicates that
\begin{eqnarray}
\frac{d p_n(\theta,\epsilon)}{d\theta} & = & \frac{1}{\epsilon}(p_{n-1}(\theta,\epsilon)-p_n(\theta,\epsilon))\\
\frac{d p_0(\theta,\epsilon)}{d\theta} & = & -\frac{1}{\epsilon}p_0(\theta,\epsilon)
\end{eqnarray}
The solution is 
\begin{equation}
p_n(\theta,\epsilon)=\frac{(\theta/\epsilon)^n}{n!}e^{-\theta/\epsilon}
\label{poisson}
\end{equation}
For obvious reasons we call this the Poisson choice. In the limit that $\epsilon\rightarrow 0$ we recover the standard Schr\"{o}dinger representation from Eq.(\ref{mas_eq}), 
\begin{equation}
\frac{d\rho(\theta)}{d\theta}=-\frac{i}{\hbar}[\hat{g},\rho(\theta)]
\end{equation}
The condition in  Eq.(\ref{Schro-limit}) is thus satisfied. We anticipate that $\nu,\ \mu$ ultimately take their values from a future quantum theory of spacetime, so here we simply equate them to the Planck time and Planck length, respectively,  in which case. $c\nu=\mu$.  We now consider the experimental consequences of this modification of the Schr\"{o}dinger rules. 

We begin with temporal translations. The differential change in a state due to a temporal translation, $t$, is given by\cite{Milburn91}
\begin{equation}
\frac{d\rho(t)}{dt}=\frac{1}{\nu}\left [e^{-i\nu\hat{H}/\hbar}\rho(t)e^{i\nu\hat{H}/\hbar}-\rho(t)\right ]
\label{milburn-eqn}
\end{equation}
where $\nu$ is a a fundamental constant with units of time. (In terms of reference \cite{Milburn91}  $\gamma=1/\nu$.  The physical consequences of this equation have been explored in \cite{Milburn91} and subsequent papers\cite {Buzek1998}.  Firstly the standard limit (Eq.(\ref{standard_time})) for the uncertainty in the estimate of a  time parameter is changed to include an additional noise source. Secondly, and most relevant for this paper, the dynamics implied by Eq.(\ref{milburn-eqn}) lead to the decay of coherence in the energy basis. 

\begin{figure}[h]
\centering
   \includegraphics[scale=0.5]{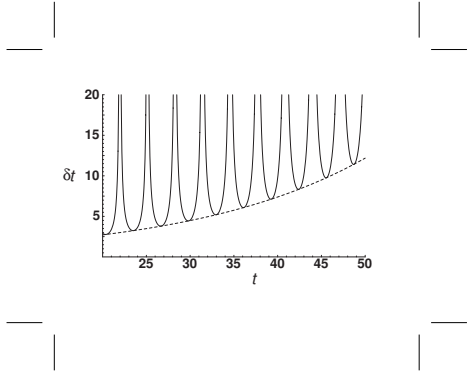}
  \caption{The modified uncertainty bound for estimating time parameters that follows from Eq.[\ref{milburn-eqn}] using the two state model in section (solid) and an approximate bound for the minima(dashed curve)}
  \label{fig1}
\end{figure}

We will use the example discussed in Section \ref{standard_time}.  The time parameter uncertainty bound   can be conveniently written in terms of the average value of the  hermitian operator $\hat{X}=|E_1\rangle\langle E_2|+|E_2\rangle\langle E_1|$ as
\begin{equation}
\sqrt{1-\langle \hat{X}(t)\rangle^2}\left | \frac{d\langle \hat{X}(t)\rangle}{dt}\right  |^{-1}
\end{equation}
The equation of motion for $\langle \hat{X}(t)\rangle$ can then be found using Eq. [\ref{milburn-eqn}]. The solution is \cite{Milburn91}
\begin{equation}
\langle \hat{X}(t)\rangle=\Re\left \{\exp\left [-\frac{t}{\nu}(1-e^{-i\nu\Delta_E})\right ]\right \}
\label{aver_X}
\end{equation}
The resulting bound on $\delta t$ is complicated, but can be simplified by the case $\nu\Delta_E<<1$,  for which we can approximate
\begin{equation}
\langle \hat{X}(t)\rangle\approx e^{-\nu\Delta_E^2 t/2}\cos(\Delta_E t)
\end{equation}
We expect that the best accuracy for the estimate of time will occur when $\langle \hat{X}(t)\rangle=0$, as at that point this moment has maximum slope. This corresponds to the condition $\cos(\Delta t)=0$. At those times we find
\begin{equation}
\delta t \approx \frac{1}{\Delta_E}e^{\nu\Delta_E^2 t/2}
\label{approx_uncer}
\end{equation}
For short times, $\Delta t<<1$ this agrees with the standard time parameter uncertainty bound for this system. However for long times we see that there  is an exponential degradation  of the accuracy.  In figure \ref{fig1} we plot  $\delta t$ versus time using the exact result in Eq.[\ref{aver_X}]. The dashed curve is the function for the  approximation given in Eq.[\ref{approx_uncer}].  This result suggests that clocks `age', that is to say long-lived atomic clocks gradually lose accuracy. Of course with $\nu$ set to the Planck time, the time  scale for this  effect is cosmological. 

The consequences for estimating a temporal translation are important as they indicate a fundamental limitation on the accuracy of clocks. This is an aspect that has been considered in some detail by Egusquiza, and  Garay \cite{EG03}, from a very different starting point.  

Now consider the case of spatial translations. Suppose we take as the fiducial state a system in a pure, minimum uncertainty, state, $|\psi_0\rangle$ with a Gaussian 
position probability density;
\begin{equation}
P_0(x)=|\langle\psi_0|x\rangle|^2=(2\pi\sigma)^{-1/2}e^{-\frac{x^2}{2\sigma}}
\end{equation}
where 
\begin{equation}
\sigma=\langle\Delta\hat{x}^2\rangle_0 =\frac{\hbar^2}{2\langle\Delta\hat{p}^2\rangle_0}
\end{equation}
where $\langle\Delta\hat{A}^2\rangle_0$ is the variance of the operator $\hat{A}$ in the fiducial state.  
Under the conventional Schr\"{o}dinger rule for displacements this state density after a displacement becomes
\begin{equation}
P^{(c)}(x|X)=(2\pi\sigma)^{-1/2}e^{-\frac{(x-X)^2}{2\sigma}}
\end{equation}
We can see that the uncertainty, $\delta X$ with which we can infer the parameter, $X$ is 
$\delta X\geq\sqrt{\sigma}/2$ or in other words $\delta X^2\langle\Delta\hat{p}^2\rangle_0\geq\hbar^2/4$, 
which is the standard result for a parameter based uncertainty principle for position\cite{BCM96}. 

In the modified Schr\"{o}dinger rule this uncertainty 
principle is modified as the width of the position  distribution is no longer independent of 
the displacement but increases linearly with displacement. The change in the state due to the displacement is given by,
\begin{equation}
\frac{d\rho(X)}{dX}=\frac{1}{\mu}\left (e^{-i\mu\hat{p}/\hbar}\rho(X)e^{i\mu\hat{p}/\hbar}-1\right)
\label{translation}
\end{equation}
This equation appears to bear a superficial relation to the recent proposal of Shalyt-Margolin and Suarez\cite{S-MS}

To see this how the uncertainty principle is changed we use Eq.(\ref{translation}) to find an equation for rate of change of the mean position and variance with displacement. It is easy to see that
\begin{eqnarray*}
\frac{d\langle\hat{x}\rangle}{dX} & = & 1\\
\frac{d\langle\hat{x^2}\rangle}{dX} & = & 2\langle\hat{x}\rangle+\mu
\end{eqnarray*}
Thus for the example here with the chosen fiducial state 
\begin{equation}
\langle\Delta\hat{x}^2\rangle_X=\langle\Delta\hat{x}^2\rangle_0+\mu X
\end{equation}
The uncertainty with which we can estimate the parameter now becomes
\begin{equation}
\delta X^2\geq \langle\Delta\hat{x}^2\rangle_0+X\mu
\end{equation}
which implies the uncertainty principle
\begin{equation}
\delta X^2\langle\Delta\hat{p}^2\rangle_0\geq\frac{\hbar^2}{4}+X\mu\langle\Delta\hat{p}^2\rangle_0
\end{equation}
This kind of modified uncertainty principle has been suggested in the context of quantum gravity and string theory\cite{Kempf}. We see here it arises as a natural consequence of an intrinsic uncertainty of spatial translations. 

We turn form intrinsic noise to the complementary process of intrinsic decoherence. In quantum mechanics noise is necessarily accompanied by decoherence. Thus any model that introduces an intrinsic uncertainty due to space tim fluctuations must necessarily introduce intrinsic decoherence. In the case of temporal translations, the decoherence occurs in the energy basis as is easily seen by computing the change in the off diagonal elements of the state in the energy basis. From Eq.(\ref{milburn-eqn}) we see that
\begin{equation}
\frac{d\rho_{i,j}(t)}{dt}=\frac{1}{\nu}\left (e^{-i\nu(E_i-E_j)/\hbar}-1\right )\rho_{i,j}(t)
\end{equation}
where $\rho_{i,j}(t)=\langle E_i|\rho(t)|E_j\rangle$ with $|E_i\rangle$ an energy eigenstate. 
This equation was discussed extensively in \cite{Milburn91}. To see the effect of intrinsic decoherence we expand the right hand side to first order in $\nu$,
\begin{equation}
\frac{d\rho_{i,j}(t)}{dt}=-i\frac{(E_i-E_j)}{\hbar}\rho_{i,j}(t)-\frac{\nu(E_i-E_j)^2}{2\hbar^2}\rho_{i,j}(t)
\end{equation}
The last term indices a decay of off diagonal coherence in the energy basis at a rate that increases quadratically with distance away from the diagonal. 

In the case of spatial translations we find a similar equation that causes a decay with, respect to the translation parameter,  of off diagonal coherence in the momentum basis,
\begin{equation}
\frac{d\rho_{k,k'}(t)}{dX}=\frac{1}{\mu}\left (e^{-i\mu(k-k')}-1\right )\rho_{k,k'}(t)
\end{equation}
where $\rho_{k,k'}(t)=\langle \hbar k|\rho(t)|\hbar k'\rangle$ with $|\hbar k\rangle$ a momentum eigenstate.  Expansion to first order in $\mu$ gives a decay of coherence in the momentum basis as the translation parameter increases. 

\section{Lorentz invariant formulation}
In order to generalise the preceding ideas to include Lorentz invariance we must move to a field theory formulation. Space and time translations are determined by specifying the
sources and detectors for the field. We are at liberty to choose any field at all,  although in practice  the electromagnetic field is the easiest to use.   The source determines the fiducial state of the quantum field. Measurements reduce to particle detectors and space and time translation parameters are inferred by the statistics of detection events at such  detectors.  The relevant unitary translation generators are still position and momentum generators, but now constructed in the usual way from whatever quantum field we wish to use. As in the non relativistic case, spatial translations require a fiducial state with an indefinite momentum while time translations require a fiducial state with and indefinite energy. 

We will first discuss how spacetime translations are determined in the standard formulation of quantum field theory. Estimation of a space time translation in quantum parameter estimation theory was considered by Braunstein et al. \cite{BCM96}, and we now summarise that treatment.  The generator for spacetime translation is the energy-momentum 4-vector
\begin{equation}
\hat{{\bf P}}=\hat{P}^\alpha {\bf e}_\alpha=\hat{P}^0{\bf e}_0+\hat{\vec{P}}=\hat{P}^0{\bf e}_0+\hat{P}^j{\bf e}_j .
\end{equation}
The spacetime translation we seek to estimate can be written as 
\begin{equation}
{\bf X}=S{\bf n}=Sn^\alpha{\bf e}_\alpha
\end{equation}
with 
\begin{equation}
{\bf n}=n^0{\bf e}_0+\vec{n}
\end{equation}
is a space-like or time-like unit 4-vector specifying the direction of the spacetime translation and $S$ is the invariant interval that parameterizes the translation.  The rotation of the fiducial state $|\psi_0\rangle$,  in Hilbert space is then 
\begin{equation}
|\psi_S\rangle=e^{iS{\bf n}\cdot\hat{{\bf P}}/\hbar}|\psi_0\rangle
\label{translation}
\end{equation}
with 
\begin{equation}
{\bf n}\cdot\hat{{\bf P}}=\eta_{\alpha\beta}n^\alpha\hat{P}^\beta=n^\alpha\hat{P}_\alpha=-n^0\hat{P}_0+\vec{n}\cdot\hat{\vec{P}}
\end{equation}
The Minkowski metric is  $\eta_{\alpha\beta}={\rm diag}(-1,+1,+1,+1)$ ( we use units such that $c=1$). The three dimensional dot product is written $\vec{n}\cdot\hat{\vec{P}}=n^j\hat{P}^j$. Braunstein et al.\cite{BCM96} show that the parameter based uncertainty principle for estimating the spacetime translation parameter, $S$ is
\begin{equation}
\langle (\delta S)^2\rangle_S\langle ({\bf n}\cdot\Delta\hat{{\bf P}}\rangle=\langle (\delta S)^2n^\alpha n^\beta\langle \Delta\hat{P}_\alpha\Delta\hat{P}_\beta\rangle\geq\frac{\hbar^2}{4N}
\end{equation}
for $N$ trials. When ${\bf n}$ is time like this is a time-energy uncertainty relation for the observer whose 4-velocity is ${\bf n}$, and when ${\bf n}$ is space-like, this is a position-momentum uncertainty relation for an observer whose 4-velocity is orthogonal to ${\bf n}$. 

What fiducial states are appropriate for estimating a spacetime translation? We will discuss the case of space and time translations separately to parallel the discussion in the nonrelativistic case. For specificity we will assume that we are using the electromagnetic field. It should be noted that this is a special case, but will suffice to illustrate the principles of the more general situation.  The energy momentum 4-vector can be written most easily if we decompose the field into plane wave modes,
\begin{equation}
\hat{{\bf P}}=\sum_{\vec{k},\sigma}\hbar{\bf k}\hat{a}^\dagger_{\vec{k},\sigma}\hat{a}_{\vec{k},\sigma}
\end{equation}
where ${\bf k}=\omega{\bf e}_0+\vec{k}=\omega{\bf e}_0+k^j{\bf e}_j$ is a null wave 4-vector with $\omega=|\vec{k}|=k$ and the sum is over all wave 3-vectors $\vec{k}$ and polarisation $\sigma$.  The  generator for spacetime translations is thus determined by the number operator for the field modes which is the generator for phase shifts in the field. Thus determining a spacetime translation via the electromagnetic field reduces to phase parameter estimation. Optimal phase estimation is not a straightforward measurement, particularly in the multimode case\cite{phase}.  However it will suffice for our purposes to give a simple example based on photon counting. (Ultimately all field measurements reduce to counting field quanta.)  

Let us consider just two modes, with wave 4-vectors ${\bf k}_1,{\bf k}_2$, with the same polarisation. We will designate a Fock state for the mode ${\bf k}_i$ as $|n\rangle_i$.  Suppose we have a source that produces the single photon state $|1\rangle_1\otimes|0\rangle_2$, ie one photon in mode ${\bf k}_1$ and the vacuum in mode ${\bf k}_2$.  The first step is to find a unitary transformation, $U$ to give
\begin{equation}
U|1\rangle_1\otimes|0\rangle_2=\frac{1}{2}(|1\rangle_1\otimes|0\rangle_2+|0\rangle_1\otimes|1\rangle_2)
\end{equation}
If the modes have the same frequency, $\omega_1=\omega_2$, this can be performed with a simple linear optical device know as a beam splitter, but if the modes also have different frequencies we need the nonlinear optical device know as a frequency converter\cite{WallsMilb}. The state is now subjected to the unitary spacetime translation in Eq.(\ref{translation}), followed by $U^\dagger$. The final state is
\begin{equation}
|\psi_S\rangle=e^{iS\delta_+/2}\left (\cos(s\delta_-/2)|0\rangle_1\otimes|1\rangle_2+i\sin(S\delta_-/2)|1\rangle_1\otimes|0\rangle_2\right )
\end{equation}
where 
\begin{equation}
\delta_\pm={\bf n}\cdot({\bf k}_1-{\bf k}_2)
\end{equation}
A simple measurement can now be made of the photon number difference between the two modes, with results $\pm 1$ occurring with probabilities
\begin{equation}
P(+1)=1-P(-1)=\sin^2(S\delta_-/2)
\end{equation}
Sampling this distribution enables an inference of the spacetime translation parameter $S$.  Of course such a measurement is not optimal. Using many photon states, and a different kind of output measurement  it is possible to do much better\cite{Sanders1995}.  

If we seek only a space  translation (ie a ruler) then we can chose the modes to have the same frequency, but wave vectors in different directions. Such a state clearly has an indefinite 3-momentum as we found for the non relativistic case. If we seek a time translation ( ie a clock) we must chose the wave vectors to have  a different frequency, that is to say a different energy as in the non relativistic case.   

It is now straightforward to define an intrinsic decoherence model that is Lorentz invariant.  The change in the state of a quantum field as a  function of the displacement interval is 
\begin{equation}
\frac{d\rho(S)}{dS}=\frac{1}{\epsilon}\left (e^{i\epsilon{\bf n}\cdot\hat{{\bf P}}/\hbar}\rho(S) e^{-i\epsilon{\bf n}\cdot\hat{{\bf P}}/\hbar}-\rho(S)\right )
\end{equation}
Equivalently
\begin{equation}
\rho(S)=\sum_{m=0}^\infty\frac{(S/\epsilon)^m}{m!}e^{-S/\epsilon} e^{im\epsilon{\bf n}\cdot\hat{{\bf P}}/\hbar}\rho_o e^{-i m\epsilon{\bf n}\cdot\hat{{\bf P}}/\hbar}
\label{relativ_state}
\end{equation}
As the generator of spacetime translations is already explicitly Lorentz invariant, these equations are Lorentz invariant. In fact $S{\bf n}\cdot\hat{{\bf P}}$  is nothing more than the action associated with the spacetime interval. The central assumption for this relativistic model of intrinsic decoherence is that the action along some worldline can vary from trial to trial in an experimental determination of a spacetime translation (this observation suggests an equivalent formulation in terms of path integrals).   The state $\rho$ is a many particle field state and would typically be specified in the Fock basis for some mode decomposition. The specific form the intrinsic decoherence takes depends on the field under discussion through the energy momentum 4-vector. We now consider some consequences of this equation for the case of the electromagnetic field. 

The most obvious modification is to the experimentally observed dispersion relation. The dispersion relation must be determined by making phase dependent measurements on the field amplitude at different spacetime points. We have postulated that such repeated measurements are described by a density operator $\rho(S)$ rather than a pure state, and we have given a rule for how to translate this state to describe measurements made at different spacetime positions. In order to measure a field amplitude that is non zero we must specify a fiducial field state that has a non zero amplitude. We will take this to be a coherent state\cite{WallsMilb}. 

In the standard theory of the electromagnetic field,  we specify the electric field at position $\vec{x}$   by the operator
\begin{equation}
 \hat{E}(\vec{x})=i\sum_{k}\omega_k(u_k(\vec{x})a_k-u_k(\vec{x})^*a_k^\dagger)
\end{equation}
where $a_k,a_k^\dagger$ are boson annihilation and creation operators, while $u_k(\vec{x})$ are a set of orthonormal mode functions and 
Let us choose the state of the field on the spacelike hypersurface $t=0$ to be a coherent state\cite{WallsMilb} such that 
\begin{equation}
{\rm tr}[ a_k\rho] =\alpha_k
\end{equation}
This is a semiclassical state for which the field amplitude on $t=0$ is given by
\begin{equation} 
{\cal E}(\vec{x})=\sum_{k}i\omega_k(u_k(\vec{x})\alpha_k-u_k(\vec{x})^*\alpha_k^*)
\end{equation}
We now translate the field along the time-like direction $n^\alpha=(1,0,0,0)$, so that the field amplitude becomes
\begin{equation}
{\cal E}(\vec{x},\tau)={\rm tr}[ \hat{E}(\vec{x})\rho(\tau)]
\end{equation}
where $\tau$ denotes the proper time. Using Eq.(\ref{relativ_state})  for this spacetime path we have that
\begin{equation}
\rho(\tau)=\sum_{n=0}^\infty\frac{(\tau/\epsilon)}{n!}e^{-\tau/\epsilon}e^{-in\epsilon\sum_k k a_k^\dagger a_k}\rho(0)e^{in\epsilon\sum_k k a_k^\dagger a_k}
\end{equation}
The field amplitude at $\tau\neq 0$ is then determined by   
\begin{eqnarray}
{\rm tr}\left [a_k\rho(\tau)\right ] & = & \sum_{n=0}^\infty\frac{(\tau/\epsilon)}{n!}e^{-\tau/\epsilon}{\rm tr}\left [a_k e^{-in\epsilon\sum_l l a_l^\dagger a_l}\rho(0)e^{in\epsilon\sum_l l a_l^\dagger a_l}\right ]\\
   & = & \sum_{n=0}^\infty\frac{(\tau/\epsilon)}{n!}e^{-\tau/\epsilon}{\rm tr}\left [\rho(0) e^{in\epsilon k a_k^\dagger a_k}a_ke^{-in\epsilon k a_k^\dagger a_k}\right ]\\ 
  & = & \alpha_k \exp\left [\frac{\tau}{\epsilon}(e^{-i\epsilon k }-1)\right ]
\end{eqnarray}
Thus
\begin{equation}
\alpha_k(\tau)=\alpha_ke^{-i\omega(k)\tau}e^{-\gamma(k)\tau}
\end{equation}
where the observed frequency of this mode amplitude is 
\begin{equation}
\omega(k)=\frac{\sin(\epsilon k)}{\epsilon}
\end{equation}
and the amplitude decays due to intrinsic decoherence at the rate 
\begin{equation}
\gamma(k)=\frac{1}{\epsilon}(\cos(\epsilon k)-1)
\end{equation}
As $\epsilon\to 0$ we recover the standard dispersion relation with a small modification
\begin{equation}
\omega(k)=k(1-\frac{\epsilon^2 k^2}{6}+\ldots)
\end{equation}

\acknowledgements I would like to thank Adrian Kent for stimulating discussions.

\end{document}